# Synthesis of nanodiamond reinforced silver matrix nanocomposites: microstructure and mechanical properties

A. Katzensteiner, J.M. Rosalie[*], R. Pippan, A. Bachmaier

Erich Schmid Institute, Austrian Academy of Sciences, Jahnstrasse 12, A-8700 Leoben, Austria

[*]Present address: Chair of Materials Physics, Montanuniversität Leoben, Jahnstraße 12, Leoben, 8700, Austria

Corresponding author: Andrea Bachmaier, andrea.bachmaier@oeaw.ac.at

**Abstract**

Nanodiamond reinforced silver matrix nanocomposites with different nanodiamond content are produced from pure silver and nanodiamond powders by a combination of two severe plastic deformation methods, ball-milling and high-pressure torsion (HPT). The ball milling parameters are varied to improve the distribution of the nanodiamonds in the powder blends. An optimized processing route combining both processes has been developed to achieve a homogenous dispersion of small nanodiamond particles in a nanocrystalline silver matrix. The microstructure of these nanocomposites was characterized by scanning and transmission electron microscopy. The mechanical properties were characterized by microhardness measurements and tensile tests. An increasing amount of nanodiamonds (up to 5 wt% and 15 vol%, respectively) can raise the microhardness by up to 70 % and the tensile strength by up to 60 % of the values achieved in HPT deformed pure silver. The microhardness and tensile strength is about 5 times higher than microcrystalline pure silver.

## 1. Introduction

Metallic composites reinforced with carbon nanoparticles such as carbon nanotubes, graphene or nanodiamonds (NDs), exhibit excellent intrinsic mechanical and functional properties like high strength, stiffness, wear resistance and excellent thermal and electric conductivity [1–7]. NDs, also known as ultra-dispersed diamonds are produced by detonation of carbon-containing explosives and subsequent purification of the detonation soot [8]. With a size range of 4 to 6 nm in diameter, high microhardness and thermal stability, NDs can be used as reinforcements for composites to improve scratch and impact resistance as well as thermal stability and thermal conductivity. However, due to their large surface to volume ratio they show a strong predisposition to stick together and form larger agglomerates [8]. This agglomeration is detrimental for NDs in metallic composites, which makes the dissolution of the agglomerates and the homogenous dispersion of the ND nanoparticles in the matrix the major challenge in producing ND reinforced matrix composites [9]. The spherical shape of the NDs helps in this regard, as they are better dispersible both in liquid solvents and in metal matrix materials than other carbon-based materials such as carbon nanotubes with their long fiber-like shape [10].

Powder-metallurgy based processes have been used to produce composite powder blends containing ND nanoparticles using different metal matrix materials (tungsten, titanium, copper, aluminum, nickel). To obtain a bulk material, different powder consolidation methods are used (e.g. hot extrusion, hot-press sintering, spark plasma sintering) proving the general applicability of these processes for the production of ND reinforced composites [3–5,7,11–16]. However, adding larger amounts of ND nanoparticles (more than 2 vol %) to the metal matrix powders lead to agglomerated ND nanoparticles in the final microstructures. Furthermore, extensive porosity and poor mechanical properties are reported due to these ND agglomerates preventing the composite powders from sintering [13,17,18].

As alternative approach, molecular-level mixing in a solution instead of conventional solid powder mixing has been used to produce homogeneous ND nanoparticle reinforced composites [2,19]. In this process, however, a surface functionalization of the NDs during deposition is necessary and a consolidation step is needed as well.

In this study, metal matrix composites with different concentrations of NDs (up to 5 wt% and 15 vol%, respectively) as the reinforcing carbon phase were processed by a combination of two severe plastic deformation methods. Ball milling of powder blends was used to break up the ND agglomerates. Special emphasis was laid on finding the optimal milling parameters to ensure a homogenous distribution of the NDs in the powder blends. Subsequent consolidation and deformation by high-pressure torsion (HPT) was performed to achieve a fully dense bulk material as well as further grain refinement of the metal matrix. HPT is a relatively simple method to produce composites with ultra-fine grained and nanocrystalline microstructures both from bulk materials and powders [20]. Currently, HPT sample dimensions range from 8 mm up to 60 mm in diameter and thicknesses of up to 12 mm can be realized [21].

Silver (Ag) was chosen as the metal matrix material in this study. Possible application examples of bulk ND reinforced Ag matrix nanocomposites are, for example, heat sinks due to the high thermal conductivity reported for ND-reinforced metals [22], or as materials were a superior wear resistance is needed [23]. ND reinforced Ag matrix nanocomposites might be also used as materials for jewelry applications, which is also possible with other matrix materials (e.g. gold).

## 2. Materials and Methods

Pure Ag powder (-500 mesh, spherical, purity: 99.9 %) provided by Alfa Aesar and ND powder (particle size: 4-6 nm, carbon purity: >98 %) provided by Plasmachem were used as starting materials. Ball-milling was conducted in a planetary ball-mill PM 400 (Retsch) in ambient atmosphere as well as in inert Argon (Ar) atmosphere. The ball-milling was conducted with alternating intervals of 15 minutes milling and 15 minutes breaks to prevent heating of the powder. The jars were filled with powder and hardened steel balls in a weight ratio of 1:10. Either pure Ag powder or mixtures of Ag and ND powder with ND contents of 0.1, 0.5, 1, 2 and 5 wt% (0.3, 1.6, 3.2, 6.3 and 14.8 vol%) were milled.

After milling, the Ag/ND composite powders were compacted and deformed by HPT. The HPT-deformation was conducted at room temperature at a hydrostatic pressure of 5 GPa and a velocity of 0.6 revolutions per minute. The HPT process is thoroughly described in Ref. [24]. Table 1 summarizes the processing parameters of all samples investigated in this study.

*Table 1: ND content in weight percent (wt%) and volume percent (vol%), milling parameters and number of HPT revolutions for all investigated Ag/ND composite samples.*

| Sample number | ND cont. [wt%] | ND cont. [vol%] | Milling time t [h] | Milling speed [rpm] | Milling atmosphere | HPT turns n |
|---|---|---|---|---|---|---|
| 1 | 0 | 0 | - | - | - | 30 |
| 2 | 0.1 | 0.3 | 2 | 150 | air | 20 |
| 3 | 0.5 | 1.6 | 2 | 150 | air | 20 |
| 4 | 0.5 | 1.6 | 2 | 150 | air | - |
| 5 | 1 | 3.2 | 3 | 150 | air | 20 |
| 6 | 2 | 6.3 | 3 | 150 | air | 30 |
| 7 | 5 | 14.8 | 6 | 150 | air | 30 |
| 8 | 0 | 0 | 6 | 300 | Ar | 30 |
| 9 | 0.1 | 0.3 | 6 | 300 | Ar | 30 |
| 10 | 0.5 | 1.6 | 6 | 300 | Ar | 30 |
| 11 | 0.5 | 1.6 | 6 | 300 | Ar | - |
| 12 | 2 | 6.3 | 6 | 300 | air | 30 |
| 13 | 5 | 14.8 | 6 | 300 | air | 30 |

The as-deformed Ag/ND samples were cut in half and embedded in resin to measure the Vickers microhardness in the tangential direction with a Buehler Micromet 5100 using a load of 100 g ($HV_{0.1}$). The Vickers microhardness is drawn as a function of the equivalent strain

$$\varepsilon_{eq} = \frac{2 * \pi * r}{t * \sqrt{3}} * n$$

with *r* being the radius of the sample, *t* the thickness of the sample and *n* the number of HPT revolutions [24].

Scanning electron microscope (SEM) images were obtained with a SEM type LEO 1525. Transmission electron microscope (TEM) images were obtained with a Phillips CM12 operating with 120 kV and high-resolution transmission electron microscope (HRTEM) images were obtained with a JEOL 2100F equipped with a $C_S$-corrector at 200 kV.

For the tensile tests, tensile test specimens with a gauge length of 2.5 mm and a gauge diameter between 300 and 500 µm were produced from HPT-deformed samples with a high-precision grinding machine [25]. From one HPT sample, two tensile test specimens were produced and the tensile axis of each tensile specimen was set to be 2 mm from the center of the HPT disc. Tensile tests were conducted at room temperature with a Kammrath & Weiss tensile stage equipped with a 200 N load cell and a testing velocity of 2.5 µm s$^{-1}$. The grinding machine and the fabrication process as well as the tensile test setup and the evaluation software are thoroughly described in Ref. [25].

## 3. Results

3.1 Microhardness and microstructural evolution

Ag/ND composite powder blends with different ND contents were produced by ball-milling pure Ag powders adding increasing amounts of ND powders with increasing milling time, $t$, and a milling velocity of 150 rpm (table 1). The powders were then consolidated and deformed by HPT. Fig.1a shows the results of the microhardness measurements of the processed Ag/ND composite samples (correspond to samples 2-7 in table 1) as well as the microhardness of a pure Ag powder sample, which was only HPT-deformed for 30 revolutions (sample 1 in table 1). The microhardness at a radius of 3 mm ($\varepsilon \sim 465$) increases from 105 $HV_{0.1}$ for the pure HPT-deformed Ag reference sample to 160 $HV_{0.1}$ for the Ag/ND composite sample with 5 wt% NDs ball-milled for 6 hours at 150 rpm and subsequently HPT-deformed for 30 revolutions. In addition to the HPT-deformed samples, the average microhardness of an Ag 0.5 wt% ND composite sample, ball-milled for 2h at 150 rpm and compacted at 5 GPa hydrostatic pressure without further HPT-deformation (sample 4), is given (77 $HV_{0.1}$) to show the effect of subsequent HPT-deformation on the microhardness evolution of the composites.

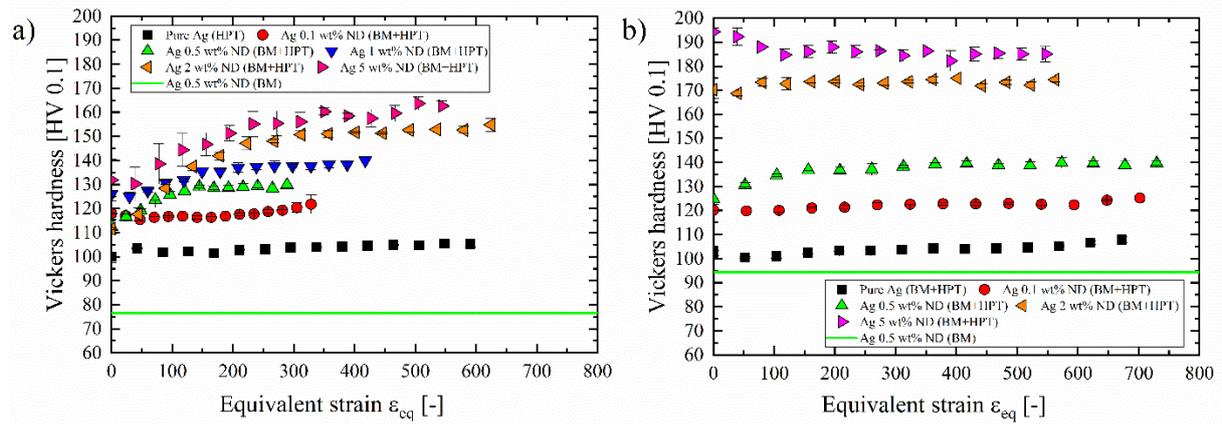

*Fig. 1: Vickers microhardness ($HV_{0.1}$) as function of the equivalent strain for pure HPT-deformed Ag powder samples and HPT-deformed Ag/ND composite samples with varying amounts of NDs: a) ball-milled with varying milling times with 150 rpm and b) ball milled for 6h with 300 rpm.*

During HPT deformation, a saturation of the grain refinement process is usually observed at a certain amount of strain. Despite the increase in milling duration (3h or 6h instead of 2h for ND contents ≥ 1 wt%), the onset of microhardness saturation (corresponds to the saturation of the grain refinement process in the Ag matrix) goes to higher values of equivalent strain with increasing ND content (Fig.1a). To ensure saturation of microhardness over the whole sample even for high ND contents, the ball-milling velocity was increased from 150 rpm to 300 rpm, first for powder blends with 0.5 wt% ND content (sample 11 in table 1), which were compacted at 5 GPa hydrostatic pressure without further HPT-deformation. These milling parameters (6 h at 300 rpm) also result in an increased microhardness of the compacted, but not deformed composites (c.f. Fig.1a and b). Powder blends with 0.1, 0.5, 2 and 5 wt% NDs as well as pure Ag powder were then produced with the same milling parameters (samples 8-13 in table 1). While the samples containing 0.1 and 0.5 wt% NDs exhibit nearly the same microhardness compared to the with 150 rpm milled samples, the samples with higher ND content (2 and 5wt% ND) showed a remarkably increased microhardness. Furthermore, the microhardness is almost constant over the whole radius for all ND contents indicating a saturation of the microstructure (Fig.1b). However, the microhardness of the pure HPT-deformed Ag sample is nearly the same, irrespective of the ball milling speed. Overall, a microhardness increase of more than 80 $HV_{0.1}$ compared to the pure Ag reference samples could be achieved by adding NDs as reinforcement phase.

To show the development of the ND nanoparticle distribution depending on the processing parameters and the ND content, SEM micrographs were taken in tangential direction of the HPT processed disks. Fig. 2 shows micrographs taken at an equivalent strain of 250 of the pure Ag reference sample (sample 1: no milling) and for the composite samples with varying milling durations ball milled with 150 rpm. Despite a good ND distribution for composites with low ND content, huge ND agglomerates are visible in the composites containing 2 and 5 wt% NDs.

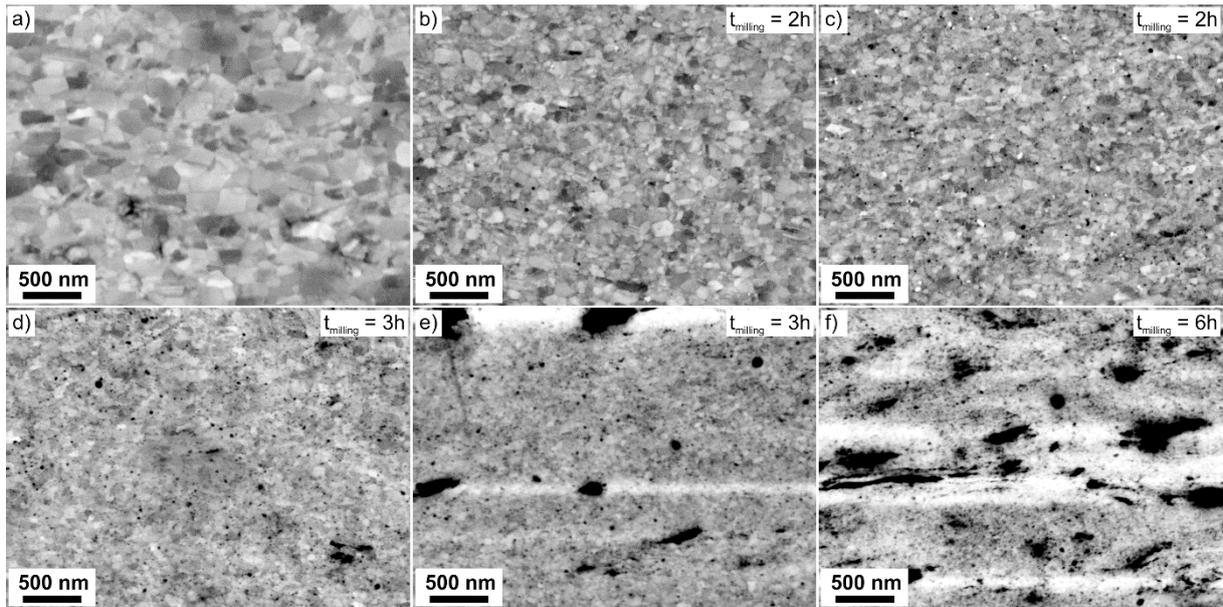

*Fig. 2: SEM micrographs at an equivalent strain of 250: a) HPT-deformed pure Ag reference sample and HPT-deformed Ag/ND composite samples, ball-milled with 150 rpm with a ND content of b) 0.1 wt%, c) 0.5 wt%, d) 1 wt%, e) 2 wt% and f) 5 wt%. The milling time is also given as inset.*

Fig.3 shows the micrographs for the samples ball-milled at 300 rpm (samples 8-13 in table 1) taken at the same equivalent strain of 250. The grain sizes of the pure Ag reference sample and the Ag/ND composite samples with 0.1 and 0.5 wt% NDs show no significant differences compared to the samples ball-milled with 150 rpm, while the samples with higher ND content have grain sizes which cannot be properly resolved by SEM for both processing conditions (150 rpm and 300 rpm, cf. Fig.2 d-f and Fig.3 d,e). The distribution of the NDs on the other hand is much better for the samples ball-milled at 300 rpm, especially for higher ND contents (cf. Fig.2 and 3 d, e).

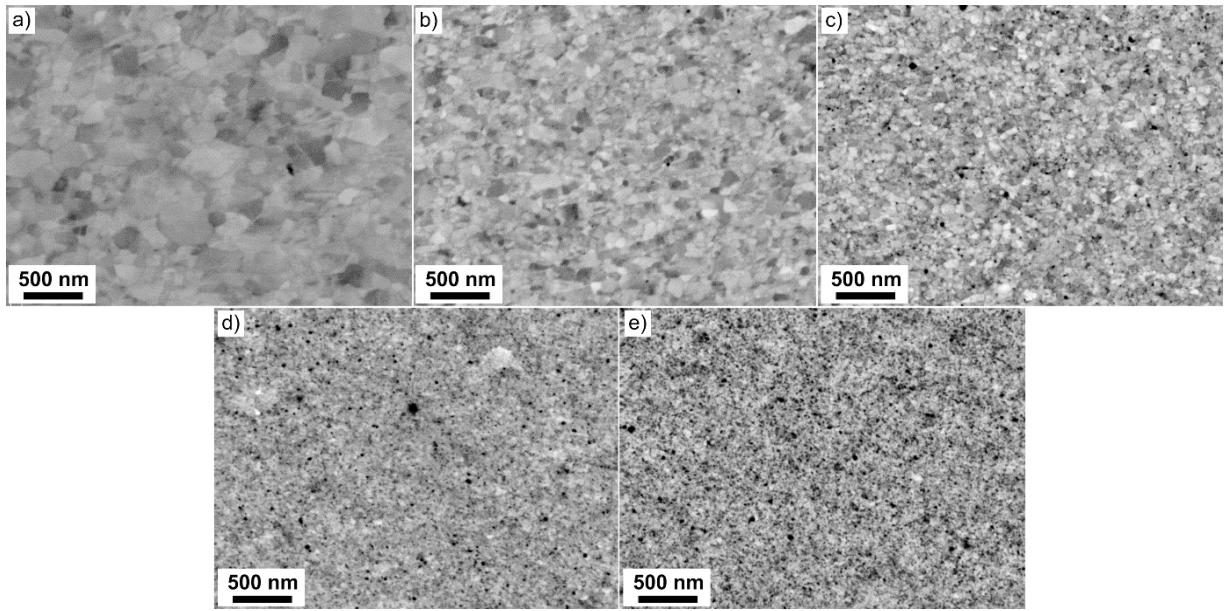

*Fig.3: SEM micrographs an equivalent strain of 250: a) HPT-deformed pure Ag powder sample and Ag/ND composite samples, ball-milled with 300 rpm for 6 hours with a ND content of b) 0.1 wt%, c) 0.5 wt%, d) 2 wt% and e) 5 wt%.*

To give an example on how the second phase distribution develops over the sample radius, SEM micrographs of the Ag/ND composite sample with 2 wt% NDs ball-milled at 300 rpm taken at a distance of 0, 1, 2 and 3 mm from the center are shown in Fig.4. In general, the ND distribution in the matrix improves with increasing strain. At low strain (Fig.4a), accumulation of larger ND agglomerates is still visible (exemplary marked by white circles). At a strain larger than 160 (Fig.4b-d), these agglomerates are separated and a more homogeneous ND distribution is achieved. The nearest neighbor distance, the aspect ratio and the size of the ND clusters were measured with the program ImageJ [26] from SEM images. While the nearest neighbor distance and the aspect ratio do not change significantly along the radius (about 35 nm and 1.5, respectively), the diameter of the ND cluster decreases slightly from 17 nm near the center to 11 nm at a sample radius of 3 mm.

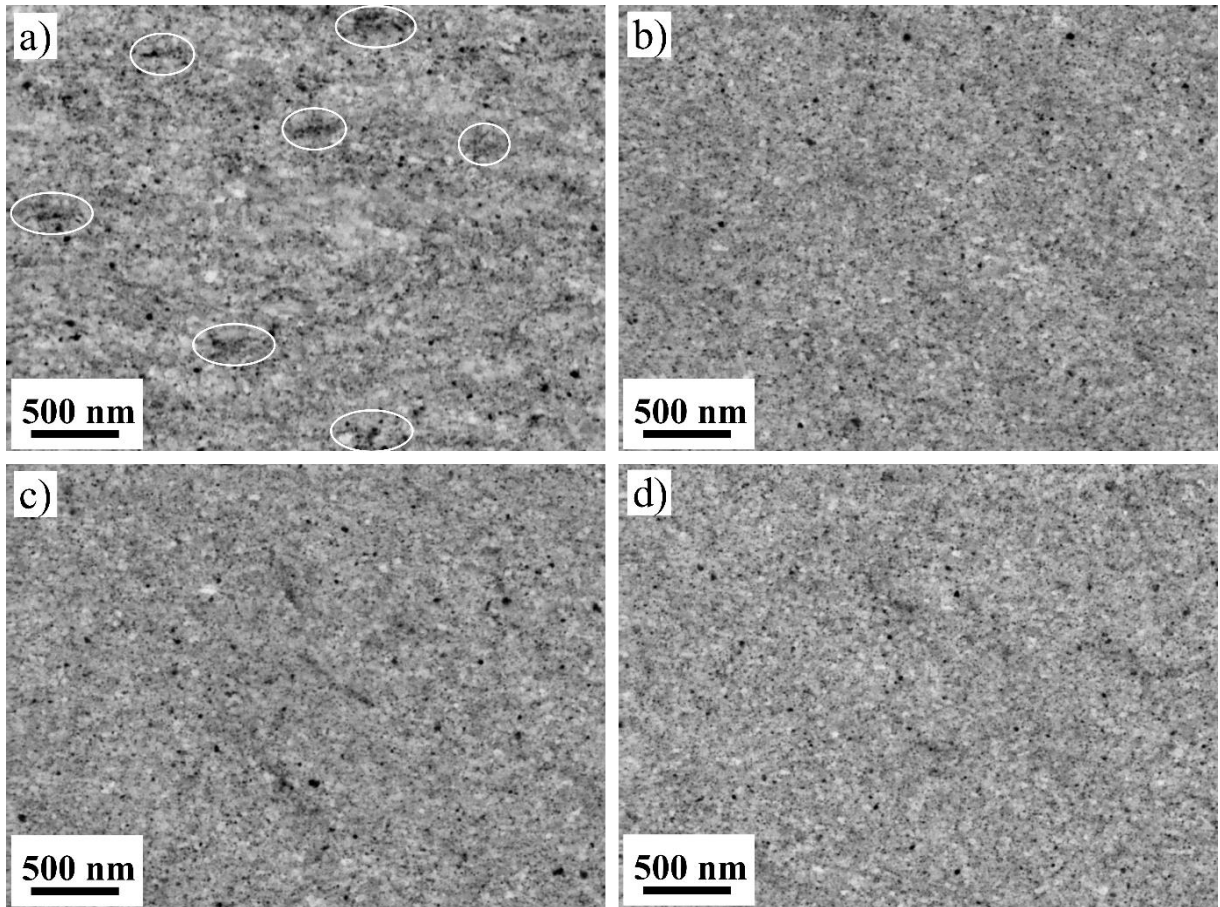

*Fig. 4: SEM micrographs of a HPT-deformed 2 wt% Ag/ND composite sample, ball-milled with 300 rpm for 6 hours, taken at a radius of a) 0 mm, b) 1 mm, c) 2 mm and d) 3 mm. Larger ND agglomerates are indicated by white circles in Fig. 4a.*

To resolve the small grain sizes of the Ag/ND composite samples with high ND content (> 0.5 wt%) as well as possible small ND clusters in the microstructure, TEM investigations were performed. Fig. 5 shows TEM micrographs and corresponding selected area electron diffraction (SAED) patterns of the composite samples with 2 wt% NDs ball-milled for 3 hours at 150 rpm (sample 6 in table 1, Fig.5a) and for 6 hours at 300 rpm (sample 12 in table 1, figure 6 b). Both microstructures consist of equiaxed, nanocrystalline Ag grains with ND clusters. The ND clusters are located at grain boundaries and triple junctions of the Ag matrix. Ag grain sizes were determined from the TEM images as the diameters of equivalent circles of around 300 grains for each sample. They were found to be 49 ± 16 nm for the sample milled with 150 rpm and 43 ± 16 nm for the sample milled with 300 rpm. Fig. 5b also includes a HRTEM image of a ND cluster located at a triple junction. The cluster might consist of several single crystal ND particles or of a single polycrystalline ND particle, indicated by ordered crystal structures (arrows in Fig.5b). The recorded SAED patterns show only Ag and no diamond diffraction peaks or peaks from possible impurities.

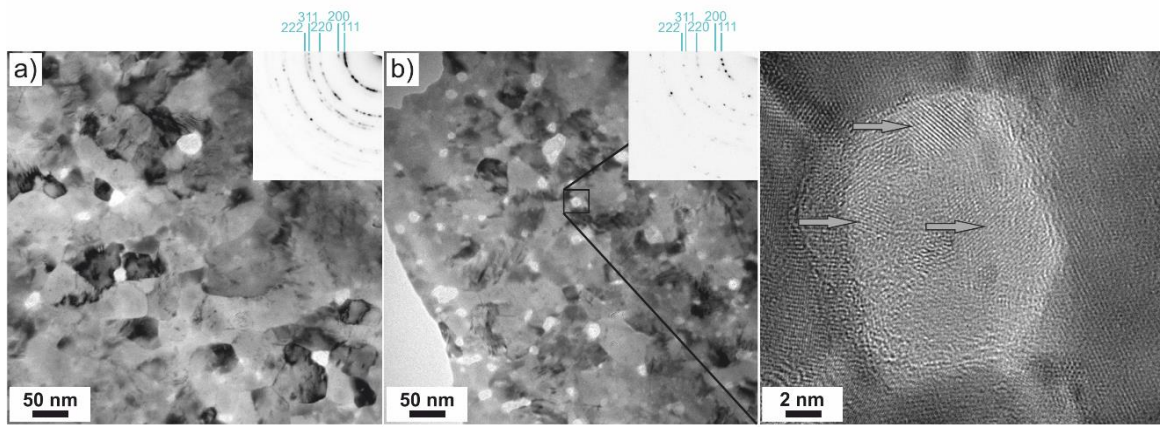

*Fig.5: TEM micrographs and SAED patterns for HPT-deformed 2 wt% Ag/ND composite samples ball-milled a) with 150 rpm and b) 300 rpm, including a HRTEM image of an ND cluster (arrows indicate ordered structures).*

3.2 Mechanical properties

To determine the influence of the different ND contents on the mechanical properties of the nanocomposites, tensile tests were performed. The tensile axis of each tensile specimen was set to be 2 mm from the center of the HPT disc. The processing parameters were chosen to be 6 hours at 300 rpm for ball-milling and 30 revolutions for the subsequent HPT-deformation. Therefore, a homogeneous distribution of the NDs in the testing region and a steady state of the microstructure of the Ag matrix is guaranteed. This is further confirmed by the constant microhardness at a strain larger than 150 as shown in Fig.1b and the microstructural evolution visible in the SEM micrographs in Fig. 4. Characteristic engineering stress-strain curves of pure HPT-deformed Ag powder, ball-milled Ag powder and ball-milled composite samples with 0.1, 0.5 and 2 wt% ND content are shown in Fig. 6. Ag/ND composite samples with 5 wt% NDs were not tested due to crack formation during HPT-deformation. In table 2, the mean values and standard deviations for the ultimate tensile strength (UTS), the yield strength, the uniform elongation, the strain at fracture and the area reduction of all investigated composite samples are summarized. UTS and yield strength are slightly higher for HPT-deformed pure ball-milled Ag samples compared to pure HPT-deformed Ag powder samples and increase further for the composite samples with increasing ND content. The uniform elongation, strain at fracture and area reduction on the other hand decrease.

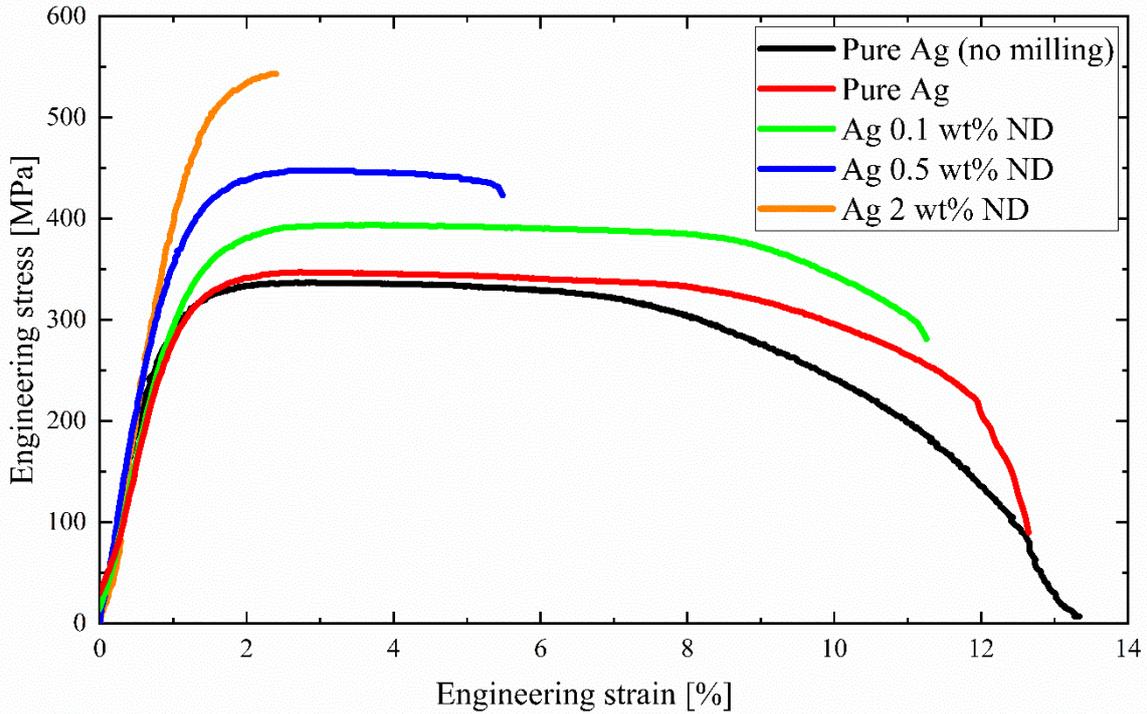

*Fig. 6: Engineering stress-strain curves of pure HPT-deformed Ag powder, ball-milled Ag powder and ball-milled composite samples with 0.1, 0.5 and 2 wt% ND content, all HPT-deformed for 30 revolutions.*

*Table 2: Mechanical properties of HPT-deformed Ag and Ag/ND composite samples.*

| HPT-def. Sample | UTS [MPa] | Yield strength [MPa] | Uniform elongation [%] | Strain at fracture [%] | Area reduction [%] |
|---|---|---|---|---|---|
| pure Ag | 329 ± 11 | 280 ± 10 | 3.0 ± 0.8 | 12.6 ± 1.2 | 94 ± 4 |
| ball-milled Ag | 350 ± 7 | 303 ± 7 | 2.2 ± 0.2 | 10.1 ± 2.1 | 74 ± 6 |
| 0.1 wt% ND | 387 ± 7 | 324 ± 4 | 2.5 ± 0.2 | 8.8 ± 1.5 | 66 ± 18 |
| 0.5 wt% ND | 438 ± 7 | 352 ± 15 | 1.9 ± 0.4 | 4.2 ± 0.6 | 21 ± 9 |
| 2 wt% ND | 531 ± 18 | 477 ± 10 | 0.9 ± 0.3 | 0.9 ± 0.3 | 4.4 ± 1.1 |

After tensile testing, fractographs were made with SEM (Fig.7). Low magnification images show a different appearance of the fracture surfaces with increasing ND content from almost complete necking of the pure Ag reference tensile specimen (Fig.7a) with an area reduction of 94 % to a more brittle failure with only a small amount of necking and an area reduction of only 4 % of the 2 wt% Ag/ND composite tensile specimen (Fig.7e). The micrographs with higher magnification show microductile fracture surfaces with dimpled structures for all tensile specimens. The dimple size is largest for the two HPT-deformed pure Ag powder tensile specimens and decrease significantly with increasing ND content. Tensile specimens made out of composite samples, which were initially ball milled before HPT deformation, further display some inclusions of different sizes inside the dimples (arrows in Fig.7).

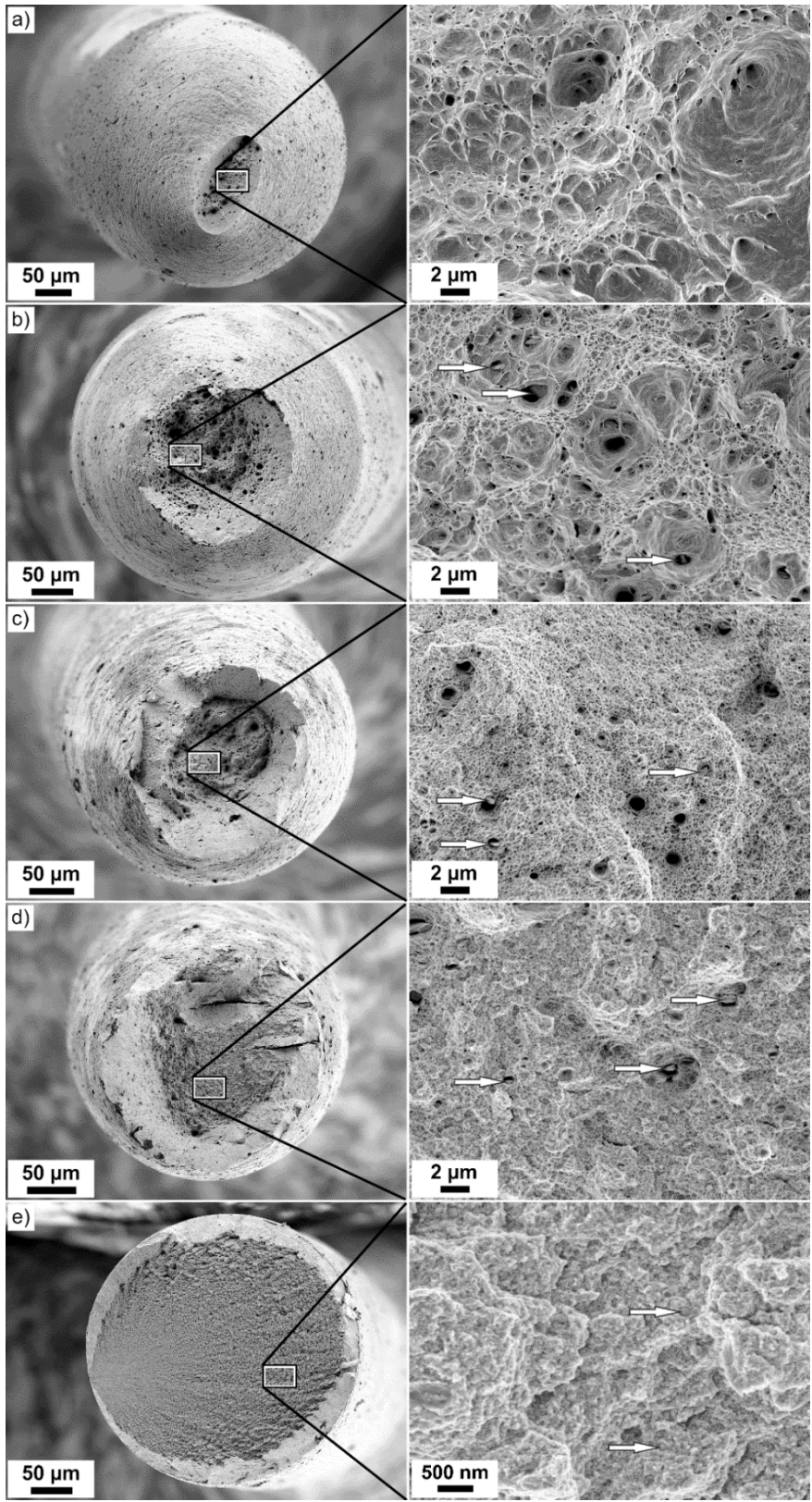

*Fig.7: SEM images of fracture surfaces of tensile specimens from a) pure HPT-deformed Ag powder, b) ball-milled Ag powder and ball-milled composite samples with c) 0.1, d) 0.5 and e) 2 wt% ND content (arrows mark inclusions).*

## 4. Discussion

*4.1 Influence of SPD on the microstructure and ND distribution*

A uniform distribution of the reinforcement phase plays a key role in providing excellent properties of ND reinforced nanocomposites [1,2,10,13]. In this study, ball milling has proven to be very effective for ND de-agglomeration. To further emphasize the effect and importance of ball milling, the microstructure of a HPT-deformed Ag/ND composite powder blend with 0.5 wt% NDs without initial ball milling is shown in Fig.8. In comparison to a previously ball-milled and HPT deformed Ag/ND composite sample with the same amount of NDs (Fig. 3c), the size of the ND agglomerates is significantly larger and still in the micrometer range.

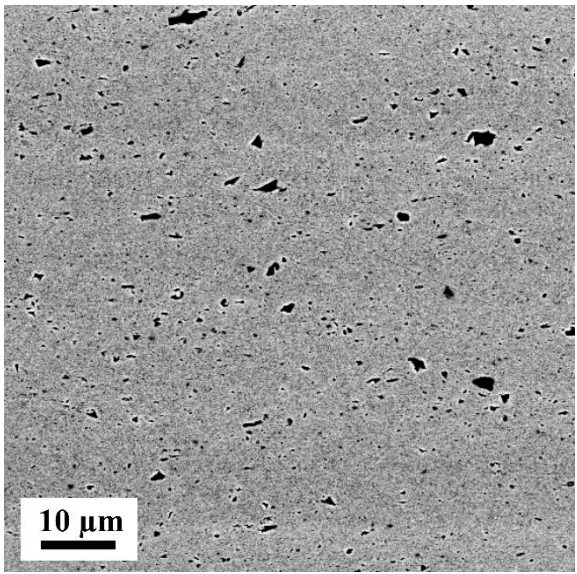

*Fig. 8: a) SEM micrograph of a 0.5 wt% Ag/ND composite sample, only HPT-deformed for 10 revolutions at RT, taken at a radius of 3 mm.*

Milling time and velocity are further important factors to achieve a break-down of large agglomerates and subsequent uniform dispersion of NDs in the metal powder, especially for higher ND contents. Changing the ball-milling velocity from 150 rpm to 300 rpm in combination with prolonged milling times results not only in smaller agglomerates, but the ND distribution is also improved (c.f. Fig. 2 and 3). Several previous studies have also shown that an increased milling time leads to a better ND distribution and consequently to a higher microhardness in ND composite powder blends [1,7,13,27]. Ball-milling on its own has, however, no large influence on the final grain size of the Ag matrix after HPT deformation. As can be seen from Fig. 2a and 3a, similar grain sizes are achieved in the pure HPT-deformed Ag reference samples with and without initial milling. Different ball milling parameters have also no significant influence on the final grain size after HPT deformation. TEM images of the 2 wt% Ag/ND composite samples ball-milled at different velocities display similar final grain sizes of the Ag matrix (49±16 nm and 43±16 nm for 150 rpm and for 300 rpm, respectively). HPT-deformation further leads to a decrease of the grain size of ball milled pure Ag with increasing equivalent strain, which is also clearly revealed in the microhardness plots in Fig. 1. Without subsequent HPT deformation, a significant smaller microhardness is achieved in the consolidated powder blends. Ball-milling and compaction of Ag powder with 0.5 wt% NDs result in a microhardness of about 77 $HV_{0.1}$ for the 150 rpm sample and 94 $HV_{0.1}$ for the 300 rpm sample. Additional HPT-deformation further increases the microhardness to nearly the same saturation hardness values (about 130 and 140 $HV_{0.1}$ for the 150 rpm and 300 rpm sample, respectively).

In pure metals, grain refinement during HPT deformation is limited due to grain boundary mobility and saturation grain sizes well above 100 nm are achieved in pure HPT deformed Ag powders [28,29]. In HPT deformed composites, the grain boundary mobility is greatly reduced by second phase particles [20]. Therefore, much smaller Ag matrix grain sizes are reached in ND reinforced Ag (Fig. 2 and 3). The influence of the HPT-deformation on the ND de-agglomeration is less prevalent than on the grain refinement. Although some initial de-clustering is observed, the size of the remaining ND clusters only decreases slightly after a certain amount of strain ($\varepsilon_{eq}$=160), and the ND distribution then shows no further improvement (Fig.4).

Furthermore, nanosized ND clusters can be found in the final microstructure after ball-milling and subsequent HPT-deformation. These ND cluster consists in similar parts of ordered diamond and possible amorphous structures (Fig.5b). In [30], HRTEM investigations of detonation NDs showed that as-synthesized NDs have crystalline structures with negligible fractions of non-diamond carbons in the initial state. The increased amount of amorphous carbon found in this study suggests that during ball-milling and HPT-deformation, the NDs might be partially damaged. In ND reinforced Al composites [1], the diamond Raman peak of ND disappeared after ball-milling due to surface graphitization of the ND particles, which maintained their diamond structure only in the particle core. A similar effect through HPT-deformation was measured by Aristizabal et al. [31] in Ni reinforced with carbon nanotubes. Here, Raman spectroscopy investigations showed a fragmentation of the carbon nanotubes into graphite/graphene clusters caused by the imposed strain.

*4.2 Impact of NDs on mechanical properties*

After optimized processing conditions were found, the mechanical properties of the HPT-deformed Ag powder samples and Ag/ND composite samples were evaluated by tensile tests. The strength is known to increase both for ball-milled and HPT-deformed metals due to the decreased grain size and a concurrent decrease of the ductility has to be accepted for most of these materials. In this study, an UTS of 329 and 350 MPa with a reduction in area of over 90 and 70% have been measured for pure Ag (without and with initial ball milling, respectively). Several studies on the mechanical properties of SPD deformed Ag have shown similar results of strength and ductility [32,33] .

The mechanical properties of metal matrix composites (MMCs) not only depend on the properties of the matrix and the reinforcement phase, but also on the structure and characteristic of the interface between the matrix and the reinforcement. A further increase in strength and decrease of ductility is observed by the additions of NDs in this study (Fig. 7 and Table 2). The UTS, for example, increases from 350 MPa to 387, 438 and 531 MPa for the Ag/ND composite samples with 0.1, 0.5 and 2 wt%, respectively. Since the same processing parameters (ball milling and HPT) have been used, it is reasonable to ascribe the observed strength increase to the presence of the NDs in the microstructure.

One major contribution to the strength results from the final grain size of the Ag matrix. The increasing amount of NDs reduces the grain boundary mobility during HPT deformation which leads to a decreasing grain size with increasing ND content of the Ag matrix (Fig. 2 and 3). The relationship between grain size and yield strength is explained by the Hall-Petch theory, which states that the strength increases proportional to $1/\sqrt{d}$ with d being the grain diameter [34]. In Fig. 9, the yield strength measured in the tensile tests is plotted as a function of the grain size. The grain sizes were measured from SEM images for the pure Ag samples and the Ag/ND composite sample with 0.1 wt% and from TEM images for the Ag/ND composite samples with 0.5 and 2 wt%. The yield strength increases with decreasing grain size following the Hall-Petch relation. Comparisons of the measured yield strength and the predicted yield strength using the Hall-Petch equation given in [34] are also in good agreement (table 3). For these calculations, the measured grain sizes and the constants of pure Ag ($\sigma_0 = 29 \pm 3$ MPa, k = 100 MPa µm$^{1/2}$ [34,35]) were used, which explains the slightly different values compared to the linear fit (c.f. table 3 and Fig.9). Thus, it is assumed that strengthening by additional grain refinement compared to the pure Ag samples is the main contributor for the observed yield strength of the Ag/ND MMCs.

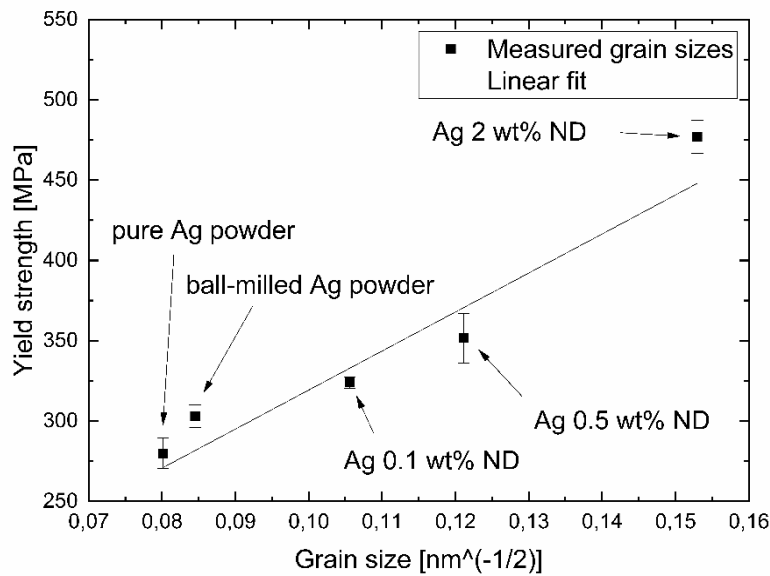

*Fig. 9 Yield strength as a function of grain size.*

Besides stabilizing finer grain sizes, introducing nanoparticles into a material can increase the strength of the composite by dispersion strengthening. If a high density of NDs would be located in the grain interior, these NDs would certainly lead to a further enhancement of the yield strength, which has been observed, for example, in nanocrystalline Fe-Zr alloys [36,37]. However, a uniform distribution of intragranular NDs is a prerequisite for dispersion strengthening. The TEM investigations, however, show that the ND particles are located at the grain boundaries and triple junctions of the Ag matrix and thus do not increase the strength of the Ag/ND composite samples by dispersion strengthening.

The strength increase with increasing volume fraction of NDs might be further correlated by a simple rule of mixture like in conventional particle reinforced composites [38]. To achieve an effective reinforcement, the ND nanoparticles should be evenly distributed throughout the matrix and their volume fraction must be high enough [38]. With a higher ball-milling velocity, and therefore a better ND distribution, an additional strength increase can be observed for all Ag/ND composite samples. Several other studies confirm a strength increase trough ND addition for metal matrix composites [1,2,7,13,27,39]. NDs as reinforcements have also been shown to result in MMCs with higher strength compared to other carbon derivatives such as carbon nanotubes or carbon onions [10].

Other studies claim that another factor in providing high strength for MMCs is the load transfer between the matrix and the second phase. A good bonding and a high aspect ratio are, however, necessary for an efficient load transfer from the softer matrix to the harder reinforcement phase. HRTEM images of Ag/ND composite samples with 2 wt% NDs indicate, that ball-milling for 6h at 300 rpm and subsequent HPT-deformation leads to small ND clusters (Fig.5). In Table 3, the modified Shear Lag model is used to predict the contribution in strengthening due to load transfer in the Ag/ND composites [40]. For particles with a low aspect ratio as in the case of NDs, the strengthening by load transfer depends mainly on the volume fraction of the reinforcement [40]. Load transfer has therefore only a small influence on the overall strength of the Ag/ND composite samples with less than 15 vol% reinforcements (Table 3).

*Table 3: Measured yield strength and calculated strength values for proposed strengthening mechanisms of HPT-deformed Ag powder samples and Ag/ND composite samples.*

| HPT-deformed Sample 30 turns | Measured Yield strength [MPa] | Hall-Petch strength component [MPa] | Load-transfer strength component [MPa] |
|---|---|---|---|
| Pure Ag powder | 280 ± 10 | 283 ± 3 | -- |
| Ag powder ball-milled | 303 ± 7 | 296 ± 3 | -- |
| 0.1 wt% ND | 324 ± 4 | 363 ± 3 | 0.77 ± 0.29 |
| 0.5 wt% ND | 352 ± 15 | 412 ± 4 | 4.17 ± 0.87 |
| 2 wt% ND | 477 ± 10 | 513 ± 4 | 17.08 ± 7.21 |

Solid solution strengthening by dissolved carbon atoms as well as the introduction of impurities during ball-milling also have to be considered when evaluating the contributing factors to the strength. A high potential for solid solution strengthening by dissolved carbon is given [41–43]. However, to achieve a significant contribution of solid solution strengthening to the overall strength, it must be assumed that all or a significant amount of carbon is actually in solid solution in Ag. This is clearly not the case in our Ag/ND composite samples as are large amount ND clusters are visible after HPT deformation. Therefore, the major portion of carbon is still bound in the NDs. As the ND clusters are dispersed during ball-milling and HPT, single carbon atoms could be removed from NDs and dissolved in the Ag matrix. Even if we would assume that a significant amount of carbon is dissolved inside the Ag grains due to deformation induced mixing as observed in Fe-C systems [44,45], segregation of carbon to grain boundaries can be expected due to the small size and low solubility of carbon in Ag [46]. Thus, the observed strength increase in our Ag/ND composite samples is mainly governed by the small structural size of the Ag matrix grains in the composites.

Hydrogen and oxygen from the atmosphere and iron from the milling jars are the main impurity elements introduced by ball milling. The milling time and velocity, the milling atmosphere as well as the microhardness difference between powder and milling medium are all important factors in this regard. SAED patterns of the 2 wt% Ag/ND composite samples ball-milled for 3 hours at 150 rpm and for 6 hours at 300 rpm show only peaks of pure Ag. Since SAED measurements only cover very small areas and are not sensitive to trace impurities, possible impurities might not be detected. The assumption that the milled Ag powder blends are contaminated by impurities during ball milling is, however, strengthened by experimental observations. First, the higher microhardness of the compacted Ag/ND composite sample with 0.5 wt% NDs ball-milled at 300 rpm compared to that ball-milled at 150 rpm (Fig.1). Second, the presence of inclusions on the fracture surface of the initially ball-milled samples not present in the case of the only HPT deformed pure Ag reference sample (sample 1).

The reduction in area and morphology of the fracture surfaces are important characteristics of the ductility of a material and depend like the other mechanical properties on the matrix grain size, the size and distribution of the NDs and on the bonding between them and the matrix. Large agglomerates of NDs have to be thoroughly broken-down and well dispersed in the Ag matrix to minimize the probability of crack initiation inside those agglomerates or at the interface between them and the matrix [2]. With increasing ND content, the reduction in area decreases. Furthermore, the dimple sizes on the fracture surface decrease. Whereas no inclusion can be found on the fracture surface of pure HPT-deformed Ag powder sample (Fig.7a), the initially ball milled pure Ag sample shows inclusions within the dimples (arrows in Fig.7b). It is assumed that these inclusions also result from impurities picked up during the ball-milling process from the milling media. The fracture surfaces of the 0.1 wt% and 0.5 wt% Ag/ND composite samples (Fig.7c and d) show mainly very small and only a few larger dimples containing about 1 μm long inclusions. The 2 wt% Ag/ND composite sample (Fig.7e) shows an almost brittle fracture surface at low magnification, but high magnification images reveal very small dimples and some small (~200 nm), more spherical inclusions. In all samples investigated, these inclusions are present in approximately half of the larger dimples visible in the micrographs and show no apparent fracture. Therefore, the fracture process is initiated by decohesion or weak interfacial bonding between the NDs and the Ag matrix, and not by fracture of the NDs [47]. The decohesion of the NDs and the Ag matrix leads to the formation of voids in the material, which then grow with increasing tensile stress. When the voids are large enough, they begin to coalesce, a crack forms and ultimately the material fractures [47]. In contrast, Hanada et al. [13] found in Cu/ND composites ball-milled and consolidated by spark plasma sintering that large ND agglomerates showed fracture during compression tests. Another processing method, hot extrusion of the same Cu/ND composites, provided a better dispersion of the NDs and no ND agglomerate fracture could be observed. A homogeneous dispersion and de-agglomeration of NDs into small clusters can therefore be seen as essential for a strong and reasonably ductile ND-MMC.

**Conclusion**

In this study, a series of ND reinforced Ag MMCs were successfully synthesized using ball milling and HPT deformation and the effects of milling parameters and the ND content on the final microstructure and mechanical properties of the as-deformed MMCs were investigated. Powder blends with homogeneously dispersed ND nanoparticles within the Ag matrix can be synthesized by optimized milling parameters. HPT deformation is less effective in ND dispersion and ND particle size reduction, but lead to a nanostructured metallic Ag matrix during powder consolidation and deformation. The MMCs produced showed homogenous dispersion of ND agglomerates and presented a remarkable strength over pure Ag with tensile strength values exceeding 500 MPa. The homogeneous dispersion of NDs and Ag matrix strengthening by grain refinement play the essential role in improving the mechanical properties of the MMCs.

**Acknowledgments**

A. Bachmaier and A. Katzensteiner gratefully acknowledge the financial support by the Austrian Science Fund (FWF): I2294-N36.

**Data availability**

The raw/processed data required to reproduce these findings cannot be shared at this time as the data also forms part of an ongoing study.